\documentclass{article}
\usepackage{spconf,amsmath,graphicx}
\usepackage{amssymb}
\usepackage[capitalize]{cleveref}
\usepackage{tabularx}
\usepackage{booktabs}
\usepackage{adjustbox}
\usepackage{multirow}
\usepackage{algorithm,algorithmic}
\usepackage{subcaption}
\usepackage{xcolor}
\DeclareMathOperator*{\argmax}{arg\,max} 

\usepackage{enumitem}
\setlist{nosep, leftmargin=14pt}

\usepackage{mwe} 

\title{Investigating the feasibility of Patch-based Inference for Generalized Diffusion Priors in Inverse Problems for Medical Images}
\name{Saikat Roy$^{\ast}$, Mahmoud Mostapha$^{\ast}$, Radu Miron$^{\dagger}$, Matt Holbrook$^{\ast}$, Mariappan Nadar$^{\ast}$}

\address{$^{\ast}$ Siemens Healthineers, Princeton, NJ, USA \qquad
    $^{\dagger}$Siemens Industry Software, Romania (Advanta)}
\newcommand{\flip}[1]{\reflectbox{\rotatebox[origin=c]{180}{#1}}}
\newcommand\blfootnote[1]{%
  \begingroup
  \renewcommand\thefootnote{}\footnote{#1}%
  \addtocounter{footnote}{-1}%
  \endgroup
}

\begin{document}
%
\maketitle
\begin{abstract}
Plug-and-play approaches to solving inverse problems such as restoration and super-resolution have recently benefited from Diffusion-based generative priors for natural as well as medical images. However, solutions often use the standard albeit computationally intensive route of training and inferring with the \textit{whole image} on the diffusion prior. While patch-based approaches to evaluating diffusion priors in plug-and-play methods have received some interest, they remain an open area of study. In this work, we explore the feasibility of the usage of patches for training and inference of a diffusion prior on MRI images. We explore the minor adaptation necessary for artifact avoidance, the performance and the efficiency of memory usage of patch-based methods as well as the adaptability of whole image training to patch-based evaluation -- evaluating across multiple plug-and-play methods, tasks and datasets.
\end{abstract}
\section{Introduction}
\blfootnote{Disclaimer: The concepts and information presented in this paper/presentation are based on research results that are not commercially available. Future commercial availability cannot be guaranteed.}

Image restoration problems involve the recovery of clean image $x$ from its noisy measurement $y = Hx+n$, where $H$ is a degradation matrix and $n$ is additive Gaussian noise of standard deviation $\sigma$. This can be reformulated as a solution $\hat{x}$ for the following optimization problem:
\begin{equation}
\hat{x} = \argmax_x \frac{1}{2}||y-Hx||^2 + \lambda \phi(x)
\label{eqn:opt_1}
\end{equation}
where $\frac{1}{2}||y-Hx||^2$ is a data-fidelity term, $\lambda$ is the trade-off parameter and $\phi(x)$ is the data prior. In \cite{Zhang_2017_CVPR}, it was demonstrated that while \cref{eqn:opt_1} could be decoupled and solved iteratively as in \cite{Geman_1995_nonlinear}, $\phi(x)$ could be reformulated as an image \textit{Denoiser}. Diffusion models have since been adopted as fast denoising priors in image restoration tasks \cite{daras2024survey, mardani2023variational}. While image restoration methods \cite{chung2023diffusion,zhu2023denoising} themselves are usually computationally independent of the type of inference in the diffusion priors, the later have primarily favored whole image inference.

In contrast, \textit{patch-based} inference in Diffusion models usually involve inference over sub-regions (or patches) of the input image, along with some form of global aggregation. However, numerous methods focus on generation while requiring specialized architectures \cite{skorokhodov2024hierarchical, ding2023patched, cho2024pd}. Fewer still deal specifically with plug-and-play approaches to inverse problems or specifically radiological data such as Magnetic Resonance Imaging (MRI). While \cite{hu2024learningimagepriorspatchbased,hu2024patchbaseddiffusionmodelsbeat} has explored the benefits of patch-based inference in CT abdominal images with positional embeddings, this area remains underexplored.

In this work, we demonstrate using a generalized Diffusion-based prior trained over a large MRI dataset of multiple anatomies that: \textbf{a)} Patch-wise trained models can offer comparable performance whole image training, \textbf{b)} Patch-based inference as in \cite{hu2024learningimagepriorspatchbased} can be generically used regardless of training scheme or plug-and-play method, \textbf{c)} Patch-based training renders prior models more pliable to patch-based plug-and-play inference, \textbf{d)} We highlight expected memory benefits of using patches but also that these benefits plateau on continuously reducing patch sizes. Overall, we offer insights for usage and feasibility of patch-based training and inference in plug-and-play techniques for medical image restoration.


\label{sec:intro}

\section{Method}
\subsection{Training of a Single Generalized Diffusion Prior}
\label{sec:diff-prior}
In contrast to prior work \cite{hu2024learningimagepriorspatchbased,askari2024bi,song2022solving, chung2023solving}, we use a single diffusion prior $p_\theta(x) \thicksim \phi(x)$ parameterized by $\theta$, trained on a large and diverse dataset of MRI images. 
Our prior is trained on a diverse data distribution populated by a large collection of approximately 289,000 MR images including brain, knee, prostate and other body regions from 1.5T, 3T scanners etc.
The advantage of training a diverse prior instead of ones specific to certain anatomies is that a single prior can be used across inverse problems in multiple anatomical regions (for example, knees, brain etc.). 
The evaluation data used for validating this and the preparation of this article, was obtained from the NYU fastMRI Initiative database \cite{knoll2020fastmri}, which is further discussed in \cref{sec:experimental_setup}. 

\subsection{Shifted-Grid Inference}
\label{sec:shifted-grid}
We adopt the patch-based inference scheme (termed \textit{Shifted-Grid} in this work), introduced by Patch-based position-aware Diffusion Inverse Solver (PaDIS) \cite{hu2024learningimagepriorspatchbased,hu2024patchbaseddiffusionmodelsbeat}. The use of a shifting grid allows the smoothing of grid-based artifacts, which would otherwise be visible if patches are naively stitched together. However, unlike \cite{hu2024learningimagepriorspatchbased}, we observe that the shifted grid inference can be independently applied to multiple inverse solvers such as DPS \cite{chung2023diffusion} or DiffPIR \cite{zhu2023denoising}. This is regardless of the use of positional embeddings as in \cite{hu2024learningimagepriorspatchbased} or adherence to a prior trained on a specific anatomical region. 




\subsection{Avoidance of Foreground-to-Background Boundary Transition Artifacts}
\label{sec:artifact}
\begin{figure}[!t]
    \centering
    \includegraphics[width=0.85\linewidth, trim={0 0 0 7cm},clip]{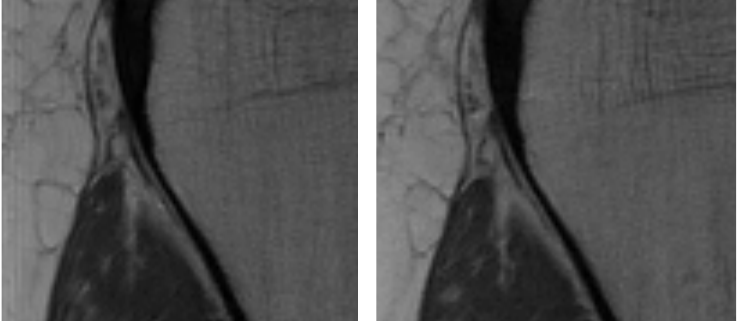}
    \caption{\textbf{Foreground-to-Background Boundary Transistion Artifacts during patchwise inference.} These artifacts are seen on the left edge of the image in the case of \textit{zero-padding} (left) while being absent for \textit{reflection padding} (right).}
    \label{fig:fg2bg}
\end{figure}

The usage of a single diverse prior trained on a large dataset with multiple anatomical regions transitions our prior, $\phi(x)$, away from a consistent idea of \textit{location}. The proximal solution in image restoration techniques such as DiffPIR or DPS struggles particularly when foreground transitions to background, leading to horizontal or vertical lines as artifacts near generated image boundaries. This is accentuated in subsequent steps by the denoising prior. However, \textit{zero-padding} is a standard part of shifted grid inference (\cref{sec:shifted-grid}) and necessarily generates such foreground-background transitions. One simple solution to avoid such artifacts is to simply avoid boundary discontinuities by switching to \textit{reflection-padding}. This results in images free of transition artifacts as illustrated in \cref{fig:fg2bg}.


\begin{table*}[ht]
    \centering
    \begin{adjustbox}{width=0.48\textwidth}
    \begin{tabular}{ccrcccc}
        \multicolumn{7}{c}{\textbf{\large Knee MRI Dataset}}\\
        \toprule
        \multirow{13}{*}{\rotatebox{90}{\textbf{Denoising}}} & \multicolumn{2}{c}{\textbf{Inference}} & \multicolumn{4}{c}{\textbf{Training Image Size}} \\ \cmidrule(lr){2-3} \cmidrule(lr){4-7}
        & \multirow{2}{*}{\textbf{Solver}} & \multirow{2}{*}{\textbf{Image Size}} & \multicolumn{2}{c}{$\mathbf{256 \times 256}$} & \multicolumn{2}{c}{$\mathbf{128 \times 128}$} \\ \cmidrule(lr){4-5} \cmidrule(lr){6-7}
         &  & & \textbf{PSNR} $\uparrow$ & \textbf{LPIPS} $\downarrow$ & \textbf{PSNR} $\uparrow$ & \textbf{LPIPS} $\downarrow$ \\
         \cmidrule{2-7}
         & \multirow{3}{*}{\textbf{DPS}} & $256 \times 256$ & $27.88_{\pm 1.09}$ & $0.19_{\pm 0.02}$ & $27.93_{\pm 1.11}$ & $0.20_{\pm 0.02}$ \\
         & & $128 \times 128$ & $27.77_{\pm 1.07}$ & $0.19_{\pm 0.03}$ & $27.86_{\pm 1.05}$ & $0.19_{\pm 0.02}$\\
         & & $64 \times 64$ & $27.67_{\pm 1.05}$ & $0.19_{\pm 0.03}$ & $27.85_{\pm 1.09}$ & $0.20_{\pm 0.02}$\\
         \cmidrule(lr){2-7}
         & \multirow{3}{*}{\textbf{DiffPIR}} & $256 \times 256$ & $27.85_{\pm 1.12}$ & $0.19_{\pm 0.03}$ & $27.75_{\pm 1.08}$ & $0.18_{\pm 0.02}$ \\
         & & $128 \times 128$ & $27.80_{\pm 1.05}$ & $0.19_{\pm 0.02}$ & $27.76_{\pm 1.07}$ & $0.18_{\pm 0.02}$ \\
         & & $64 \times 64$ & $27.68_{\pm 1.05}$ & $0.19_{\pm 0.02}$ & $27.71_{\pm 1.11}$ & $0.19_{\pm 0.02}$\\
        \bottomrule \bottomrule
    \end{tabular}
    \end{adjustbox}
    \begin{adjustbox}{width=0.48\textwidth}
    \begin{tabular}{ccrcccc}
        \multicolumn{7}{c}{\textbf{\large Brain MRI Dataset}} \\
        \toprule
        \multirow{13}{*}{\rotatebox{90}{\textbf{Denoising}}} & \multicolumn{2}{c}{\textbf{Inference}} & \multicolumn{4}{c}{\textbf{Training Image Size}} \\ \cmidrule(lr){2-3} \cmidrule(lr){4-7}
        & \multirow{2}{*}{\textbf{Solver}} & \multirow{2}{*}{\textbf{Image Size}} & \multicolumn{2}{c}{$\mathbf{256 \times 256}$} & \multicolumn{2}{c}{$\mathbf{128 \times 128}$} \\ \cmidrule(lr){4-5} \cmidrule(lr){6-7}
         &  & & \textbf{PSNR} $\uparrow$ & \textbf{LPIPS} $\downarrow$ & \textbf{PSNR} $\uparrow$ & \textbf{LPIPS} $\downarrow$ \\
         \cmidrule{2-7}
         & \multirow{3}{*}{\textbf{DPS}} & $256 \times 256$ & $28.72_{\pm 1.19}$ & $0.18_{\pm 0.03}$ & $28.64_{\pm 1.23}$ & $0.17_{\pm 0.03}$ \\
         & & $128 \times 128$ & $28.57_{\pm 1.19}$ & $0.18_{\pm 0.03}$ & $28.64_{\pm 1.27}$ & $0.16_{\pm 0.03}$ \\
         & & $64 \times 64$ & $28.35_{\pm 1.16}$ & $0.20_{\pm 0.04}$ & $28.64_{\pm 1.29}$ & $0.17_{\pm 0.03}$\\
         \cmidrule(lr){2-7}
         & \multirow{3}{*}{\textbf{DiffPIR}} & $256 \times 256$ & $28.75_{\pm 1.24}$ & $0.20_{\pm 0.04}$ & $28.52_{\pm 1.29}$ & $0.16_{\pm 0.03}$ \\
         & & $128 \times 128$ & $28.63_{\pm 1.17}$ & $0.18_{\pm 0.03}$ & $28.48_{\pm 1.26}$ & $0.16_{\pm 0.03}$\\
         & & $64 \times 64$ & $28.46_{\pm 1.19}$ & $0.19_{\pm 0.04}$ & $28.47_{\pm 1.27}$ & $0.16_{\pm 0.03}$\\
        \bottomrule \bottomrule
    \end{tabular}
    \end{adjustbox}
    \begin{adjustbox}{width=0.48\textwidth}
    \begin{tabular}{ccrcccc}
        \multirow{10}{*}{\rotatebox{90}{\textbf{Super-Resolution ($\times2$)}}} & \multirow{2}{*}{\textbf{Solver}} & \multirow{2}{*}{\textbf{Image Size}} & \multicolumn{2}{c}{$\mathbf{256 \times 256}$} & \multicolumn{2}{c}{$\mathbf{128 \times 128}$} \\ \cmidrule(lr){4-5} \cmidrule(lr){6-7}
        & & & \textbf{PSNR} $\uparrow$ & \textbf{LPIPS} $\downarrow$ & \textbf{PSNR} $\uparrow$ & \textbf{LPIPS} $\downarrow$ \\
         \cmidrule{2-7}
         & \multirow{3}{*}{\textbf{DPS}} & $256 \times 256$ & $27.87_{\pm 1.75}$ & $0.25_{\pm 0.03}$ & $27.87_{\pm 1.78}$ & $0.26_{\pm 0.03}$ \\
         & & $128 \times 128$ & $27.72_{\pm 1.71}$ & $0.26_{\pm 0.03}$ & $27.75_{\pm 1.71}$ & $0.27_{\pm 0.03}$ \\
         & & $64 \times 64$ & $27.66_{\pm 1.71}$ & $0.26_{\pm 0.03}$ & $27.75_{\pm 1.72}$ & $0.27_{\pm 0.03}$ \\
         \cmidrule(lr){2-7}
         & \multirow{3}{*}{\textbf{DiffPIR}} & $256 \times 256$ & $27.86_{\pm 1.77}$ & $0.25_{\pm 0.03}$ & $27.84_{\pm 1.77}$ & $0.27_{\pm 0.04}$ \\
         & & $128 \times 128$ & $27.76_{\pm 1.71}$ & $0.26_{\pm 0.03}$ & $27.77_{\pm 1.72}$ & $0.27_{\pm 0.03}$ \\
         & & $64 \times 64$ & $27.68_{\pm 1.70}$ & $0.26_{\pm 0.03}$ & $27.77_{\pm 1.75}$ & $0.27_{\pm 0.03}$ \\
        \bottomrule
    \end{tabular}
    \end{adjustbox}
    \begin{adjustbox}{width=0.48\textwidth}
    \begin{tabular}{ccrcccc}
        \multirow{10}{*}{\rotatebox{90}{\textbf{Super-Resolution ($\times 2$)}}} & \multirow{2}{*}{\textbf{Solver}} & \multirow{2}{*}{\textbf{Image Size}} & \multicolumn{2}{c}{$\mathbf{256 \times 256}$} & \multicolumn{2}{c}{$\mathbf{128 \times 128}$} \\ \cmidrule(lr){4-5} \cmidrule(lr){6-7}
        & & & \textbf{PSNR} $\uparrow$ & \textbf{LPIPS} $\downarrow$ & \textbf{PSNR} $\uparrow$ & \textbf{LPIPS} $\downarrow$ \\
         \cmidrule{2-7}
         & \multirow{3}{*}{\textbf{DPS}} & $256 \times 256$ & $28.91_{\pm 1.35}$ & $0.16_{\pm 0.02}$ & $28.93_{\pm 1.32}$ & $0.17_{\pm 0.02}$ \\
         & & $128 \times 128$ & $28.70_{\pm 1.43}$ & $0.17_{\pm 0.02}$ & $28.83_{\pm 1.32}$ & $0.17_{\pm 0.02}$ \\
         & & $64 \times 64$ & $28.57_{\pm 1.49}$ & $0.17_{\pm 0.02}$ & $28.83_{\pm 1.34}$ & $0.17_{\pm 0.02}$ \\
         \cmidrule(lr){2-7}
         & \multirow{3}{*}{\textbf{DiffPIR}} & $256 \times 256$ & $28.92_{\pm 1.36}$ & $0.16_{\pm 0.02}$ & $28.98_{\pm 1.33}$ & $0.17_{\pm 0.02}$ \\
         & & $128 \times 128$ & $28.76_{\pm 1.44}$ & $0.17_{\pm 0.02}$ & $28.90_{\pm 1.34}$ & $0.17_{\pm 0.02}$ \\
         & & $64 \times 64$ & $28.61_{\pm 1.51}$ & $0.17_{\pm 0.02}$ & $28.87_{\pm 1.33}$ & $0.17_{\pm 0.02}$ \\
        \bottomrule
    \end{tabular}
    \end{adjustbox}    
    \caption{\textbf{Patch-based Models offer comparable performance to whole image models.} We demonstrate similar performance between our priors across multiple problems and plug-and-play methods and discuss nuances in \cref{sec:discussion}.}
    \label{tab:main_results}
\end{table*}

\section{Experimental Setup}
\label{sec:experimental_setup}
We use EDM2 \cite{karras2022elucidating} as the backbone architecture of our diffusion priors. We train with whole images ($256\times256$), as well as with randomly sampled patches of size $128\times128$ as in Patch Diffusion \cite{wang2024patch}. We also do not train with additional padding as in PaDIS, to evaluate the feasibility of patch-wise inference in \textit{generically} trained models. Peak Signal-to-Noise Ratio (PSNR) and Learned Perceptual Image Patch Similarity (LPIPS) \cite{zhang2018unreasonable} are used as our metrics to measure performance.
We use two inverse problems to evaluate the effectiveness of our trained diffusion priors -- Denoising and Super-Resolution ($2\times$). We use both whole image ($256\times256$) inference and at patch sizes of size $64\times64$ and $128\times128$. As mentioned in \cref{sec:diff-prior}, we use a large and diverse training set of 289,000 complex MR images.
A single slice from 62 Knee and 200 Brain volumes (not included in the training data) from the NYU fastMRI Initiative database \cite{zbontar2018fastmri,knoll2020fastmri} are used for evaluation. DPS \cite{chung2023diffusion} and DiffPIR \cite{zhu2023denoising} serve as plug-and-play approaches to solving our inverse problems.


\section{Results and Discussion}
\label{sec:discussion}

\begin{figure*}[!ht]
    \centering
    \begin{subfigure}{0.16\textwidth}
        \flip{\includegraphics[width=2.85cm, height=2.85cm]{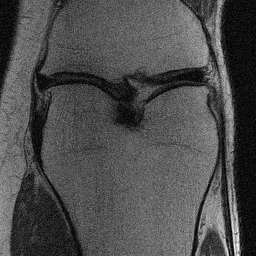}}
        \flip{\includegraphics[width=2.85cm, height=2.85cm]{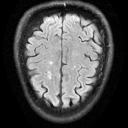}}
        \subcaption*{Noisy measurement $y$}
    \end{subfigure}
    \begin{subfigure}{0.16\textwidth}
        \flip{\includegraphics[width=2.85cm, height=2.85cm]{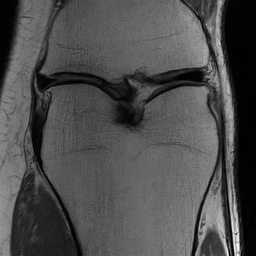}}
        \flip{\includegraphics[width=2.85cm, height=2.85cm]{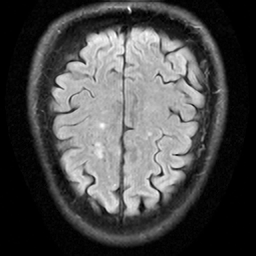}}
        \subcaption*{DiffPIR$_\text{patch}$ (64)}
    \end{subfigure}
    \begin{subfigure}{0.16\textwidth}
        \flip{\includegraphics[width=2.85cm, height=2.85cm]{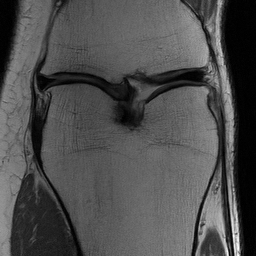}}
        \flip{\includegraphics[width=2.85cm, height=2.85cm]{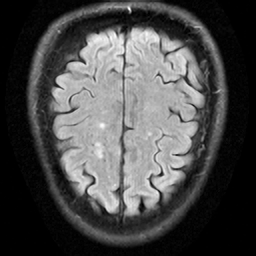}}
        \subcaption*{DiffPIR$_\text{patch}$ (128)}
    \end{subfigure}
    \begin{subfigure}{0.16\textwidth}
        \flip{\includegraphics[width=2.85cm, height=2.85cm]{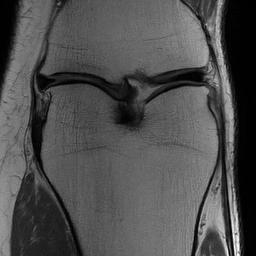}}
        \flip{\includegraphics[width=2.85cm, height=2.85cm]{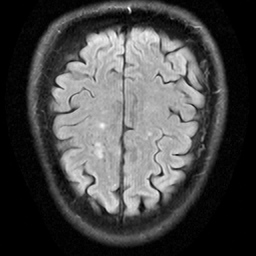}}
        \subcaption*{DiffPIR$_\text{patch}$ (256)}
    \end{subfigure}
    \begin{subfigure}{0.16\textwidth}
        \flip{\includegraphics[width=2.85cm, height=2.85cm]{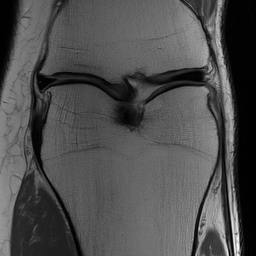}}
        \flip{\includegraphics[width=2.85cm, height=2.85cm]{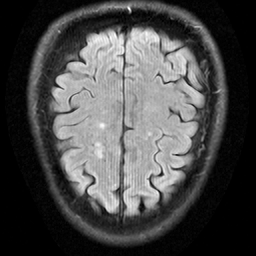}}
        \subcaption*{DiffPIR$_\text{full}$ (256)}
    \end{subfigure}
    \begin{subfigure}{0.16\textwidth}
        \flip{\includegraphics[width=2.85cm, height=2.85cm]{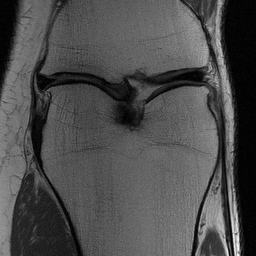}}
        \flip{\includegraphics[width=2.85cm, height=2.85cm]{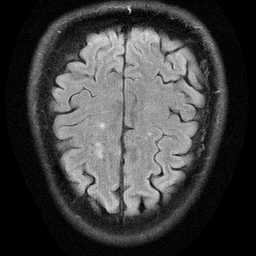}}
        \subcaption*{Clean image $x$}
    \end{subfigure}
    \caption{\textbf{Results on Denoising (Top row) and Super-Resolution (Bottom row) tasks using DiffPIR.} We demonstrate perceptually comparable performance of our diverse prior in both patch-based (DiffPIR$_\text{full}$) and patch-based DiffPIR$_\text{patch}$ training. 
    }
    \label{fig:denoising}
\end{figure*}

\subsection{Patch-wise trained models offer comparable performance to whole image models}
As seen in \cref{tab:main_results}, we demonstrate that our patch-wise  trained prior offers comparable performance to a model trained at full image size. Specifically, when our model trained with patch size $128 \times 128$ are used in whole image mode for plug-and-play, their performance is comparable to our whole image trained model, although the later is still slightly better. This is illustrated further in \cref{fig:denoising}. This highlights that patch-based training could be an effective alternative when medical image resolutions grow (for example, ultra high resolution CTs) to render whole image training memory-intensive or even infeasible.

\subsection{Generalized Patch-based inference}
We also demonstrate that regardless of the inverse problem or training method, shifted-grid based inference as discussed in \cref{sec:shifted-grid}, can be substituted into either of our plug-and-play techniques, as long as considerations are made to reduce artifacts like those at boundaries (\cref{sec:artifact}). We hypothesize the effectiveness of patch-wise evaluation without any added positional guidance \cite{hu2024learningimagepriorspatchbased} is possible due to the sense of global context offered by the proximal solution in either of our inverse solvers. This is specifically true when the prior's sense of \textit{location} is occluded by the presence of data from multiple anatomical regions in our training dataset. However, we do believe that positional information might enhance the process in cases where this is not the case.

\subsection{Patch-wise plug-and-play prefers patchwise training}
It is noticeable in \cref{tab:main_results}, that while patch-based inference can be applied regardless of training, models trained with randomly sampled patches are slightly more resilient to changing patch sizes during inference. On the other hand, both our patch-based and whole image network is seen to perform slightly better in most settings when also evaluated with whole images. This highlights a trade-off between memory usage and accuracy, both when using patches for training or inference of a diffusion prior, in plug-and-play settings.

\subsection{Patch-wise inference offers efficient memory usage}
\begin{figure}[!ht]
    \centering
    \includegraphics[width=\linewidth, height=5cm]{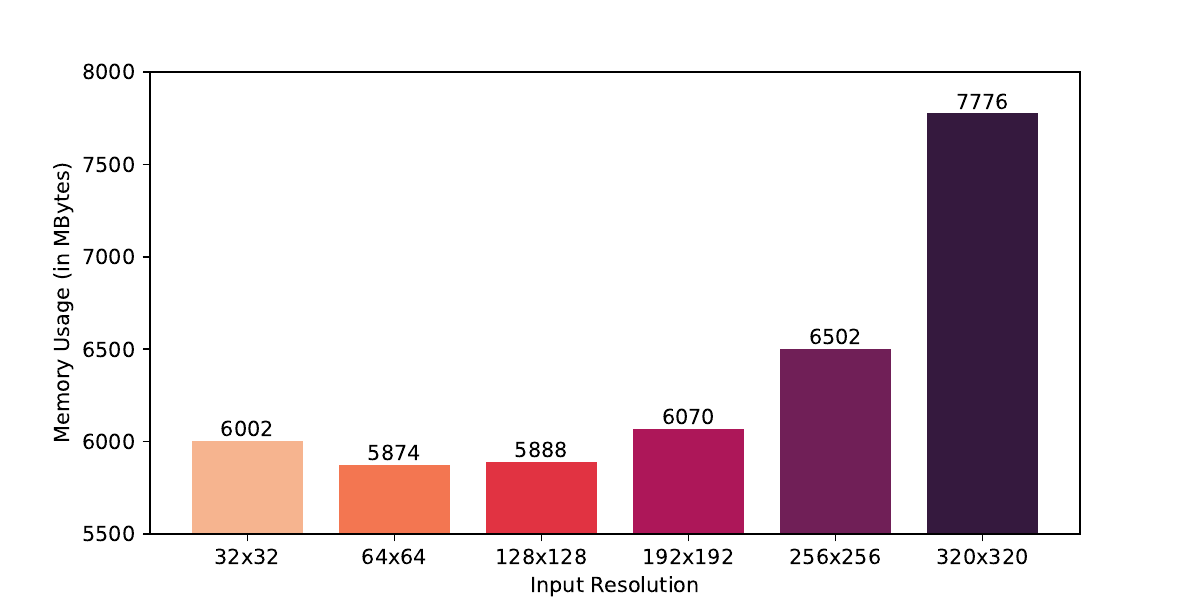}
    \caption{\textbf{Patch-based inference leads to noticeable memory reduction during inference.} This is as much as 25\% when moving from $320 \times 320$ to $128 \times 128$ patches in DiffPIR.}
    \label{fig:memory_util}
\end{figure}

As anticipated, patch-wise inference allows memory-efficient inference in plug-and-play problems. We visualize memory usage in \cref{fig:memory_util} for denoising using DiffPIR. While memory savings are noticeable, it depends on our definition of \textit{maximum image size}. For instance, we see about 25\% reduction in memory usage for patch size $128\times128$ compared to our highest evaluated size of $320\times 320$. With larger benefits for higher whole image sizes, we also see a plateau of memory usage on further reducing patch sizes to $32\times 32$ -- highlighting a limit to such gains in increasingly smaller image sizes.

\section{Conclusion}
In conclusion, we offer a deeper look into the under-explored area of patch-usage for plug-and-play techniques for medical image restoration. In doing so, we seek to enable further study into the usage on patch-based methods in memory intensive inverse problems in higher resolution medical images.

\bibliographystyle{IEEEbib}
\bibliography{strings,refs}


\end{document}